# HealthAdvisor: Recommendation System for Wearable Technologies enabling Proactive Health Monitoring

**Shubhi Asthana, Ray Strong, and Aly Megahed**
IBM Research – Almaden, San Jose, CA, USA
{sasthan, hrstrong, aly.megahed}@us.ibm.com

## Abstract

Proactive monitoring of one's health could avoid serious diseases as well as better maintain the individual's well-being. In today's IoT world, there has been numerous wearable technological devices to monitor/measure different health attributes. However, with that increasing number of attributes and wearables, it becomes unclear to the individual which ones they should be using. The aim of this paper is to provide a recommendation engine for personalized recommended wearables for any given individual. The way the engine works is through first identifying the diseases that this person is at risk of, given his/her attributes and medical history. We built a machine learning classification model for this task. Second, these diseases are mapped to the attributes that need to be measured in order to monitor such diseases. Third, we map these measurements to the appropriate wearable technologies. This is done via a textual analytics model that we developed that uses available information of different wearables to map the aforementioned measurements to these wearables. The output can be used to recommend the wearables to individuals as well as provide a feedback to wearable developers for common measurements that do not have corresponding wearables today.

## 1   Introduction

One of the most common ways today to continuously monitor the health of ever-growing population is through the use of various technologies including Internet of Things (IoT) sensors, wearable devices, smartphone, software applications, etc. However, a major problem today is the large variety of wearable devices available on the market today. According to [1], it is expected that about 148 million health monitoring devices would be shipped annually by 2019. Bonato et al. [2] discusses the future for several companies investing aggressively in development of wearable products for clinical applications. As a result, getting the most economically viable and comprehensive set of wearable technologies personalized for an individual is getting significantly challenging.

This paper aims to improve the mechanisms to provide health care in a more personalized way by recommending a selected set of wearable technologies. Our system composes of three main steps. First, we developed a machine learning classification model that maps the individual's attributes and medical history to diseases that he might be at risk of. Second, we map these diseases to the measurements that should be monitored for his case. Lastly, we developed a textual analytics model that takes the available manual of wearables in order to map different measurements to wearables so that we come up with the recommended wearables for that individual. This can then be used to achieve proactive lifestyle adjustments. Additionally, for the case that no wearables exist, our system could also trigger a count for such measurement so that such feedback can be given to technological developers later on to develop these on-demand non-existent wearable technologies.

The rest of this paper is organized as follows: in Section 2, we describe the design methodology of our recommendation model - HealthAdvisor based on the demographics information such as age, gender, location of residence, ethnicity. In Section 3 we explain our input dataset and illustrate how the results of our system look like. We then discuss the related work in Section 4 and then give the conclusions of the paper in Section 5.

## 2   HealthAdvisor System

In this section, we illustrate the three steps of our system in the three sub-sections below.

### 2.1   Identifying At-Risk Diseases for Any Given Individual

The first step in the mechanics of HealthAdvisor is a machine learning classifier that we train on the following features that we found to accurately predict the diseases that person is at risk for: the demographics information such as age, gender, location of residence, ethnicity, etc., as well as the person's Electronic Medical Records (EMR). The classifier could then be used to predict the at-risk diseases for any given person. We illustrate in Section 3 below a glimpse of our real-world data results.

From the results we have seen, for example, in a cosmopolitan city where the demographics point to high stress and pollution levels, our classification model could recommend measurements such as blood pressure and respiratory tests to identify lung disorders.

Fig 1 shows the classification model constructed using the training dataset explained in Section 4 below. It illustrates the demographic attributes of a person and his health risks based on location, age, gender. We use a Classifier algorithm to divide the dataset into different categories based on the age-group, gender, prior health records, ethnicity, demographic state and city, occupation, and marital status. The order of demographic attributes $A_i$ is selected dynamically by the algorithm to maximize the information gain G which is computed as follows: Entropy $E[D] = - \Sigma_j P(c_j) \log_2 P(c_j)$ and Information Gain $G(D, A_i) = E[D] - EA_i[D]$. $P(c_j)$ is the probability of an element belonging to class $c_j$ in the dataset D.

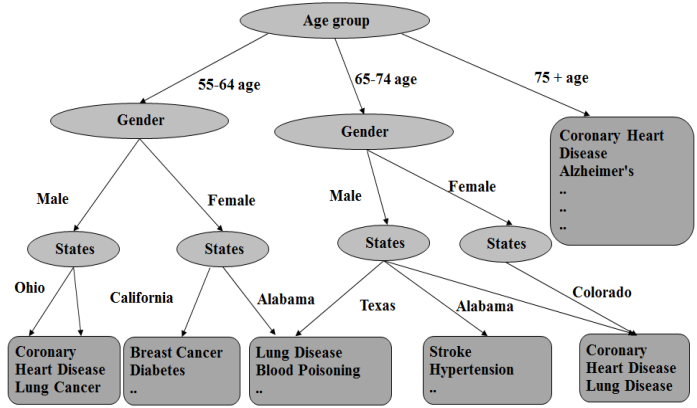

Figure 1: Classification model for prediction

The Attribute $A_i$ that has the maximum Information Gain G for a given tree level is used to split the current tree and minimize the uncertainty to partition the dataset into different classes at that level. For example, the Attribute Value "Coronary Heart Disease" is the major cause of health issue in people with Age > 75. Hence, it has the maximum Information Gain for the tree branch with attribute selection of age-group 75 and above.

We then use the demographic attributes of the person and run the model to evaluate the major causes of health issues for him. The model gives us the following vector of causes of health issues $y_i$ and corresponding probabilities $p_i$: Y : (y1 : p1 , y2 : p2 , y3 : p3 , ….. , yn : pn). We infer the personalized health track of the person that predicts the top health conditions given by Y.

### 2.2   Mapping the Predicted Diseases to Measurements

We use textual analytics on a corpus of medical data from health-care providers and clinicians to construct an entity-relationship graph between the disease risks and the Measurements. The rules used to construct this graph extract concepts as follows: Cause → Disorder → Symptoms → Measurements. Fig 2 and 3 show an example of a person for whom HealthAdvisor constructs the graph to measure tremors or imbalanced posture.

### 2.3 Mapping Measurements to Wearable Solutions

We further use Textual Analytics on a publicly available database [13] from wearable technology manufacturers to find similar measurements and extend the graph with wearable devices. Fig 3 shows the end-to-end analytics pipeline to extract the concepts including the health condition, disorder, symptoms, measurements and wearable technologies.

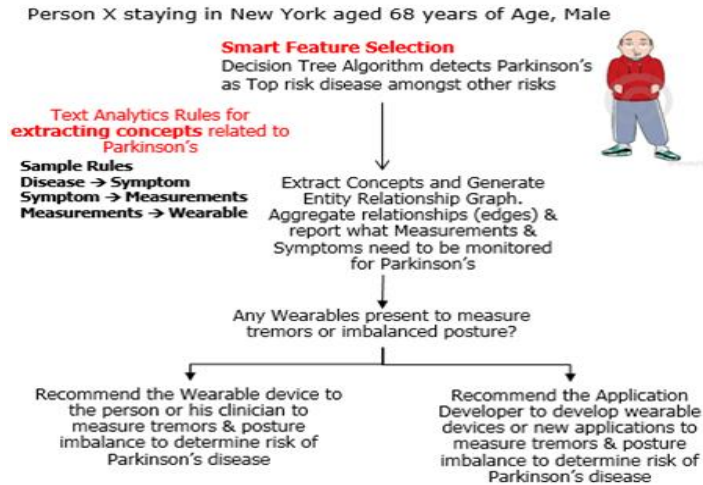

Figure 2: Textual analytics execution to generate an entity relation graph

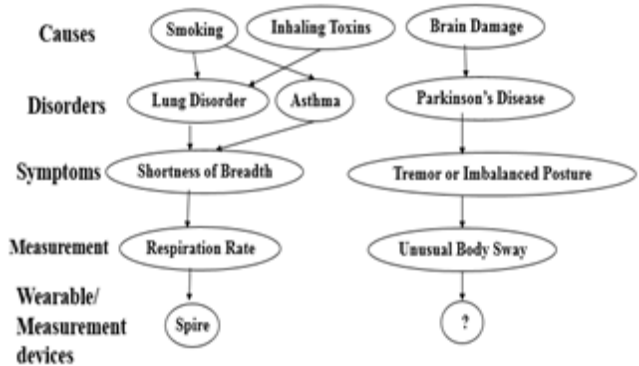

Figure 3: Entity relation graph with recommended wearable devices.

HealthAdvisor uses this rule on a Person Y staying in New York, Male having a history of smoking faces to classify for Lung disease. The textual analytic rules further help to extract concepts related to Lung disease and create the entity relationship graph. It checks for wearables using the graph and recommends the Spire [10] to test his Respiration rate.

## 3 Evaluation

We obtained the following attributes from a publicly available data source [4] and evaluated it using the Weka library [11] for different models such as Decision Tree, Logistic Regression, LibSVM and OneR [12]. We ran it for 50 target classes (disease risks) and trained it for 135,000 data points [1]. We found that the Decision Tree Model had the lowest Root Mean Square Error (RMSE) of 0.1066 and highest accuracy for all diseases. Decision Tree model also gave us a good diagrammatic representation to show which attributes of demographics help in determining the likely health risks as shown in Fig 1. Table 1 shows a glimpse of the dataset of demographic attributes and the likely at-risk diseases (called health conditions) to monitor.

Table 1: Demographic dataset with selected features

| Person | Age Group | Gender | Ethnicity | State | Occupation | Prior Health Records | Likely Health Conditions to monitor |
|--------|-----------|--------|-----------|-------|------------|----------------------|-------------------------------------|
| X | 45-54 | M | Hispanic | California | Software Engineer | Smoking | Lung Disease |
| Y | 55-64 | M | African American | Alabama | Teacher | Traffic Accident | Shock in body |
| Z | 35-44 | M | Asian | New York | Banker | Back Pain | Arthritis, Lumbar Back disease |
| P | 75+ | F | Latino | Louisiana | Retired | Instability in body | Parkinson's disease, Falls |
| Q | 75+ | M | Asian | Texas | Retired | Fracture in leg | Falls |

The Health conditions are then mapped to the Wearable Technologies or services that can help the end user in Proactive monitoring of disease as indicated in the previous sections. Table 2 shows an example output of our system for some given individuals.

Table 2: Mapping Health Conditions to Wearable Technologies

| Person | Likely Health Conditions to monitor | Wearable recommended |
|--------|-------------------------------------|----------------------|
| X | Lung Disease, Asthma | Spire, Preventice BodyGuardian |
| Y | Shock in body, Stroke | BodyTel |
| Z | Arthritis, Lumbar Back disease | Valedo |
| P | Parkinson's disease, Falls | BalanSens, LifeCall |
| Q | Falls | Sensus Pain Management Systems |

## 4      Related Work

Genetic studies have been used in the past to predict and identify the top risk diseases for a person [5, 6, and 7]. However in this paper we specifically drive insights from the demographic information of a person based on smart feature selection. In addition, we also provide an end-to-end pipeline to diagnose and recommend the most accurate and economically viable wearable technology for a personalized health care solution. Recent work on using machine learning algorithms for predicting individual disease risk has been investigated [8, 14, 15, and 16]. However, unlike HealthAdvisor, it does not recommend the wearable technologies or provide feedback to the application developer for developing unavailable health care solutions.

## 5      Conclusions

In this paper, we addressed a pressing problem for proactively finding the most likely risk factors for health conditions of individuals. We formed a recommendation system for wearable technologies for an individual. We used a classifier algorithm to build a highly trained model using a large dataset that accurately predicted the most likely health conditions that can develop in a person with a given set of demographics and a given medical history. In addition, our system uses text analytics to recommend the most relevant measurements and personalized wearable technologies for the identified likely health conditions.

**Acknowledgement**

The authors would like to thank Taiga Nakamura at IBM Research – Almaden for his helpful comments and feedback on the initial drafts of the paper.